\documentclass[twocolumn,showpacs,pre,amssymb]{revtex4}

\usepackage{graphicx}
\usepackage{amsmath}
\begin{document}

\title{Optimized suppression of coherent noise from seismic data 
using the Karhunen-Lo\`eve transform}

\author{Ra\'ul ~Montagne}
\email{montagne@df.ufpe.br}
\affiliation{Laborat\'orio de F\'\i sica Te\'orica e Computacional, 
Departamento de F\'{\i}sica, Universidade Federal de Pernambuco, 50670-901 
Recife, PE, Brazil}

\author{Giovani L.~Vasconcelos}
\email{giovani@lftc.ufpe.br}
\affiliation{Laborat\'orio de F\'\i sica Te\'orica e Computacional, 
Departamento de F\'{\i}sica, Universidade Federal de Pernambuco,
50670-901 Recife, PE, Brazil}

\begin{abstract}
Signals obtained in land seismic surveys are usually contaminated with
coherent noise, among which the ground roll (Rayleigh surface waves)
is of major concern for it can severely degrade the quality of the
information obtained from the seismic record. Properly suppressing the
ground roll from seismic data is not only of great practical
importance but also remains a scientific challenge.  Here we propose
an optimized filter based on the Karhunen--Lo\'eve transform for
processing seismic data contaminated with ground roll.  In our method,
the contaminated region of the seismic record, to be processed by the
filter, is selected in such way so as to correspond to the maximum of
a properly defined coherence index.  The main advantages of the method
are that the ground roll is suppressed with negligible distortion of
the remanent reflection signals and that the filtering can be
performed on the computer in a largely unsupervised manner. The method has been devised to filter seismic data, however  it
could also be relevant for other applications where localized coherent
structures, embedded in a complex spatiotemporal dynamics, need to be identified in a more refined way.

\end{abstract}

%Use showkeys class option if keyword
%geophysical signal processing
\maketitle

\section{Introduction}
\label{sec:intro}

Locating oil reservoirs that are economically viable is one of the
main problems in the petroleum industry. This task is primarily
undertaken through seismic exploration, where explosive sources
generate seismic waves whose reflections at the different geological
layers are recorded at the ground or sea level by acoustic sensors
(geophones or hydrophones). These seismic signals, which are later
processed to reveal information about possible oil occurrences, are
often contaminated by noise and properly cleaning the data is
therefore of paramount importance \cite{yilmaz87}. In particular, the
design of efficient filters to suppress noise that shows coherence in
space and time (and often appears stronger in magnitude than the
desired signal) remains a scientific challenge for which novel
concepts and methods are required.  In addition, the filtering tools
developed to treat such kind of noise may also find relevant
applications in other physical problems where coherent structures evolving in a complex spatiotemporal dynamics need to identified properly.

In land seismic surveys, the seismic sources generate various type of
surface waves which are regarded as noise since they do not contain
information from the deeper subsurface.  This so-called coherent noise
represents a serious hurdle in the processing of the seismic data
since it may overwhelm the reflection signal, thus severely degrading
the quality of the information that can be obtained from the data.  A
source-generated noise of particular concern is the ground roll, which
is the main type of coherent noise in land seismic records and is
commonly much stronger in amplitude than the reflected signals.
Ground roll are surface waves whose vertical components are
Rayleigh-type dispersive waves, with low frequency and low phase and
group velocities.

\begin{figure}[t]
\begin{center}
%\vspace{-0.7cm}
\resizebox{60mm}{70mm}{ \includegraphics{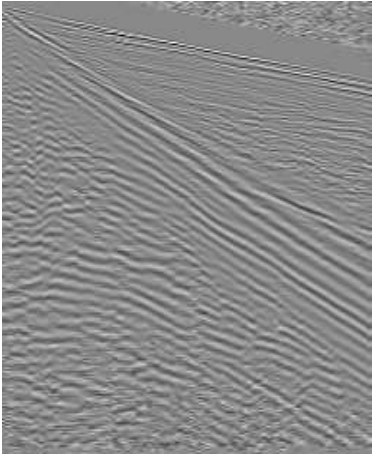} }
\end{center}
%\vspace{-1.5cm}
\caption{A space-time plot of seismic data.
The horizontal axis represents the offset distance and the vertical
axis indicates time. The origin is at the upper-left corner, and the
maximum offset and time are 475 m and 1000 ms, respectively.  The gray
scale is such that black (white) corresponds to the minimum (maximum)
amplitude of the seismic signal. The ground roll noise appears as
downward oblique lines.}
\label{fig:sismo}
\end{figure}

\begin{figure}[t]
\begin{center}
%\vspace{-0.7cm}
 \includegraphics[width=0.98\columnwidth]{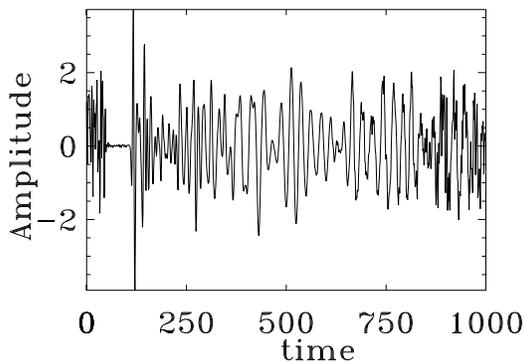} 
\end{center}
%\vspace{-1.5cm}
\caption{Seismic signal recorded by a single geophone (trace 58).
The amplitude is in arbitrary units and time in ms.}
\label{fig:trazo}
\end{figure}

An example of seismic data contaminated by ground roll is shown in
Fig.~\ref{fig:sismo}.  This seismic section consists of land--based
data with 96 traces (one for each geophone) and 1001 samples per
trace. A typical trace is shown in Fig.~\ref{fig:trazo} corresponding
to geophone 58. The image shown in Fig.~\ref{fig:sismo} was created
from the 96 traces using a standard imaging technique.  The horizontal
axis in this figure corresponds to the offset distance between source
and receiver and the vertical axis represents time, with the origin
located at the upper--left corner. The maximum offset is 475 m (the
distance between geophones being 5 m) and the maximum time is 1000 ms.
The gray levels in Fig.~1 change linearly from black to white as the
amplitude of the seismic signal varies from minimum to maximum.  Owing
to its dispersive nature, the ground roll appears in a seismic image
as a characteristic fan-like structure, which is clearly visible in
Fig.~\ref{fig:sismo}. The data shown in this figure was provided by
the Brazilian Petroleum Company (PETROBRAS).

Standard methods for suppressing ground roll include
one-dimensional high--pass filtering and two-dimensional $f$--$k$
filtering \cite{yilmaz87}. Such ``global'' filters are based on the
elimination of specific frequencies and have the disadvantage that
they also affect the uncontaminated part of the signal. Recently,
``local'' filters for suppressing the ground roll have been proposed
using the Karhunen-Lo\`eve transform \cite{liu99,tyapkin2004} and the
wavelet transform \cite{deighan97,liacir03}. The Wiener-Levinson
algorithm has also been applied to extract the ground roll
\cite{karsli2004}.

Filters based on the Karhunen--Lo\`eve (KL) transform are particularly
interesting because of the {\it adaptativity} of the KL expansion,
meaning that the original signal is decomposed in a basis that is
obtained directly from the empirical data, unlike Fourier and wavelet
transforms which use prescribed basis functions.  The KL transform is
a mathematical procedure (also known as proper orthogonal
decomposition, empirical orthogonal function decomposition, principal
component analysis, and singular value decomposition) whereby any
complicated data set can be optimally decomposed into a finite, and
often small, number of modes (called proper orthogonal modes,
empirical orthogonal functions, principal components or eigenimages)
which are obtained from the eigenvectors of the data autocorrelation
matrix. In applying the KL transform to suppress the ground roll, one
must first map the contaminated region of the seismic record into a
horizontal rectangular region. This transformed region is then
decomposed with the KL transform and the first few principal
components are removed to extract the coherent noise, after which the
filtered data is inversely mapped back into the original seismic
section. The advantage of this method is that the noise is suppressed
with negligible distortion of the reflection signals, for only the
data within the selected region is actually processed by the filter.
Earlier versions of the KL filter
\cite{liu99,tyapkin2004} have however one serious drawback, namely,
the fact that the region to be filtered must be picked by hand---a
procedure that not only can be labor intensive but also relies on good
judgment of the person performing the filtering.

In this article we propose a significant improvement of the KL
filtering method, in which the region to be filtered is selected
automatically as an optimization procedure. We introduce a novel
quantity, namely, the coherence index $CI$, which gives a measure of
the amount of energy contained in the most coherent modes for a given
selected region. The optimal region is then chosen as that that
gives the maximum $CI$. We emphasize that introducing a quantitative
criterion for selecting the `best' region to be filtered has the
considerable advantage of yielding a largely unsupervised scheme for
demarcating and efficiently suppressing the ground roll.

Although our main motivation here concerns the suppression of coherent
noise in seismic data, we should like to remark that our method may be
applicable to other problems where coherent structures embedded in a complex spatiotemporal dynamics need to be
identified or characterized in a more refined way.  For example, the
KL transform has been recently used to identify and extract spatial
features from a complex spatiotemporal evolution  in  combustion experiment
\cite{palacios98,robbins04,palacios05}. A related method---the
so-called biorthogonal decomposition---has also been applied to characterize spatiotemporal chaos and identified structures\cite{horacio04,gabo02} as well as
identify changes in the  dynamical complexity, and the spatial coherence
of a multimode laser \cite{papoff04}. We thus envision that our
optimized KL filter may find applications in these and
related problems of coherent structures in complex spatiotemporal dynamics.

The article is organized as follows. In Sec.~\ref{sec:klt} we define
the Karhunen--Lo\`eve transform, describe its main properties, and
discuss its relation to the singular value decomposition of matrices. In
Sec.~\ref{sec:KL} we present the KL filter and a novel optimization
procedure to select the noise-contaminated region to be parsed through
the filter.  The results of our optimized filter when applied to the
data shown in Fig.~\ref{fig:sismo} are presented in
Sec.~\ref{sec:results}. Our main conclusions are summarized in
Sec.~\ref{sec:conclu}. In Appendixes A and B we briefly discuss, for
completeness, the relation between the KL transform and two other
similar procedures known as proper orthogonal decomposition (or
empirical orthogonal function expansion) and principal component
analysis.

\section{The Karhunen--Lo\`eve Transform}
\label{sec:klt}

\subsection{Definition and main properties}
\label{subsec:def}

Consider a multichannel seismic data consisting of $m$ traces with
$n$ samples per trace represented by a $m\times n$ matrix $A$, so that
the element $A_{ij}$ of the data matrix corresponds to the amplitude
registered at the $i$th geophone at time $j$. For definiteness, let us
assume that $m<n$, as is usually the case. We also assume for
simplicity that the matrix $A$ has full rank, i.e., $r=m$, where $r$
denotes the rank of $A$.  Letting the vectors $\vec{x}_i$ and
$\vec{y}_j$ denote the elements of the $i$th row and the $j$th column
of $A$, respectively, we can write
\begin{equation}
A= \left(\vec{y}_1 \ \vec{y}_2 \ ... \
\vec{y}_n\right)=\left(\begin{array}{c} \vec{x}_1  \cr \vec{x}_2  \cr ... \cr
\vec{x}_m\end{array}\right). \label{eq:A}
\end{equation}
With the above
notation we have
\begin{equation}
A_{ij}=x_{ij}=y_{ji},
\end{equation}
where $a_{ij}$ denotes the $j$th element of the vector $\vec{a}_i$.
(To avoid risk of confusion matrix elements will always be denoted by
capital letters, so that a small-cap symbol with two subscripts
indicates vector elements.)

Next consider the following $m\times m$ symmetric matrix
\begin{equation}
\Gamma\equiv AA^t, \label{eg:G}
\end{equation}
where the superscript $t$ denotes matrix transposition.  It is a well
known fact from linear algebra that matrices of the form (\ref{eg:G}),
also called covariance matrices, are positive definite
\footnote{This is true if $r={\rm min}\{m,n\}$, as assumed; if
$r<{\rm min}\{m,n\}$, the matrix $\Gamma$ has $r$ nonzero eigenvalues
with all $m-r$ remaining eigenvalues equal to zero.}.  Let us then
arrange the eigenvalues $\lambda_i$ of $\Gamma$ in non-ascending
order, i.e., $\lambda_1\ge\lambda_2\ge...\ge\lambda_m>0$, and let
$\vec{u}_i$ be the corresponding (normalized) eigenvectors. 

The Karhunen-Lo\`eve (KL) transform  of the data matrix $A$ is defined as
the $m\times n$ matrix $\Psi$ given by
\begin{equation}
\Psi=U^tA, \label{eq:psi}
\end{equation}
where the columns of the matrix $U$ are
the eigenvectors of $\Gamma$:
\begin{equation}
U= \left(\vec{u}_1 \ \vec{u}_2 \ ... \ \vec{u}_m \right).\label{eq:U}
\end{equation}
The original data can be recovered from the KL transform
$\Psi$ by the inverse relation
\begin{equation}
A=U \Psi . \label{eq:AU}
\end{equation}
We refer to this equation as the KL {\it expansion} of the data matrix
$A$. To render such an expansion more explicit let us denote by the
$\vec{\chi}_i$, $i=1,...,m$, the elements of the $i$th row of the KL
matrix $\Psi$, that is,
\begin{equation}
\Psi=\left(\begin{array}{c} \vec{\chi}_1  \cr \vec{\chi}_2  \cr ... \cr
\vec{\chi}_m\end{array}\right). \label{eq:psixi}
\end{equation}
Then (\ref{eq:AU}) can be written as 
\begin{equation}
A=\sum_{i=1}^m \vec{u}_i \vec{\chi}_i , \label{eq:uxi}
\end{equation}
where it is implied matrix multiplication between the column vector
$\vec{u}_i$ and the row vector $\vec{\chi}_i$.  The eigenvectors
$\vec{u}_i$ are called {\it empirical eigenvectors}, {\it proper
orthogonal modes}, or {\it KL modes}.

As discussed in Appendix~\ref{sec:pod}, the total energy $E$ of the
data can be defined as the sum of all eigenvalues,
\begin{equation}
E=\sum_{i=i}^m\lambda_i , \label{eq:E}
\end{equation}
so that $\lambda_i$ can be interpreted as the energy captured by
the $i$th empirical eigenvector $\vec{u}_i$.  We thus define the
relative energy $E_i$ in the $i$th KL mode as
\begin{equation}
E_i=\frac{\lambda_i}{\sum_{i=i}^m\lambda_i} . \label{eq:Ei}
\end{equation}
We note furthermore that since $\Gamma$ is a covariance-like matrix
its eigenvalues $\lambda_i$ can also be interpreted as the variance of
the respective principal component $\vec{u}_i$; see
Appendix~\ref{sec:pca} for more details on this interpretation. We
thus say that the higher $\lambda_i$ the more coherent the KL mode
$\vec{u}_i$ is.  In this context, the most energetic modes are
identified with the most coherent ones and vice-versa.

An important property of the KL expansion is that it is `optimal' in
the following sense: if we form the matrix $\Psi_k$ by keeping the
first $k$ rows of $\Psi$ and setting the remaining $m-k$ rows to zero,
then the matrix $A_k$ given by
\begin{equation}
A_k=U \Psi_k \label{eq:Ak}
\end{equation}
is the best approximation to $A$ by a matrix of rank $k<m$ in the
Frobenius norm (the square root of the sum of the squares of all
matrix elements) \cite{holmes96}. This optimality property of the KL
expansion lies at the heart of its applications in data compression
\cite{andrews67} and dimensionality reduction \cite{holmes96},
for it allows to approximate the original data $A$ by a smaller matrix
$A_k$ with minimum loss of information (in the above sense).  Another
interpretation of relation (\ref{eq:Ak}) is that it gives a low-lass
filter \cite{ulrych88}, for in this case only the first $k$ KL modes
are retained in the filtered data $A_k$.

On the other hand, if the relevant signal in the application at hand
is contaminated with coherent noise, as is the case of the ground roll
in seismic data, one can use the KL transform to remove efficiently
such noise by constructing a high-pass filter.  Indeed, if we form the
matrix $\Psi'_k$ by setting to zero the first $k$ rows of $\Psi$ and
keeping the remaining ones, then the matrix ${A'}_k$ given by
\begin{equation}
A'_k=U  \Psi'_k \label{eq:Atilk}
\end{equation}
is a filtered version of $A$ where the first $k$ `most coherent' modes
have been removed.  However, if the noise is localized in space and
time  it is best to apply the filter only to the contaminated part
of the signal. In previous versions of the KL filter the choice of the
region to be parsed through the filter was made {\it a priori},
according to the best judgement of the person carrying out the
filtering, thus lending a considerable degree of subjectivity to the
process.  In the next section, we will show how one can use the KL
expansion to implement an automated filter where the undesirable
coherent structure can be `optimally' identified and removed.

Before going into that, however, we shall briefly discuss below an
important connection between the KL transform and an analogous
mathematical procedure known as the singular value decomposition of
matrices. Readers already knowledgeable about the equivalence between
these two formalisms (or more interested in the specific application
of the KL transform to filter coherent noise) may skip the remainder
of this section without loss of continuity.

\subsection{Relation to Singular Value Decomposition}
\label{subsec:svd}

We recall that the singular value decomposition (SVD) of any $m\times
n$ matrix $A$, with $m<n$, is given by the following expression:
\begin{equation}
A= U\Sigma V^t, \label{svdA}
\end{equation}
where $U$ is as defined in (\ref{eq:U}), $\Sigma$ is a $m\times m$
diagonal matrix with elements $\sigma_i=\sqrt{\lambda_i}$, the
so-called {\it singular values} of $A$, and $V$ is a $m\times n$
matrix whose columns correspond to the $m$ eigenvectors
$\{\vec{v}_i\}$ of the matrix $A^{t}A$ with nonzero eigenvalues. The
SVD allows us to rewrite the matrix $A$ as a sum of matrices of
unitary rank:
\begin{equation}
A=\sum_{i=1}^m \sigma_i Q_i=\sum_{i=1}^m \sigma_i \vec{u}_i \vec{v}_i^{~t}.
\label{eq:Q}
\end{equation}
In the context of image processing the matrices $Q_i$ are called
\textsl{eigenimages} \cite{andrews}. 

Now, comparing  (\ref{eq:AU}) with (\ref{svdA}) we  see that the KL
transform $\Psi$ is related to the SVD matrices $\Sigma$ and $V$ by
the following relation
\begin{equation}
\Psi=\Sigma V^t, \label{eq:Psi'}
\end{equation}
so that the row vectors $\vec{\chi}_i$ of  $\Psi$ are given
in terms of the singular values $\sigma_i$ and the vectors $\vec{v}_i$ by
\begin{equation}
\vec{\chi}_i=\sigma_i \vec{v}_i^{~t}.
\end{equation}
It thus follows that the decomposition in eigenimages seen in (\ref{eq:Q})
is  precisely the KL expansion given in (\ref{eq:uxi}).
Furthermore the approximation $A_k$ given in (\ref{eq:Ak}) can be
written in terms of eigenimages as
\begin{equation}
A_k=\sum_{i=1}^k \sigma_i Q_i.
\label{svdQk}
\end{equation} 
Similarly, the filtered data $A'_k$ shown in (\ref{eq:Atilk})
reads in terms of eigenimages:
\begin{equation}
A'_k=\sum_{i=k+1}^m \sigma_i Q_i.
\label{svdQtilk}
\end{equation} 
The SVD provides an efficient way to compute the KL transform, and we
shall use this method in the numerical procedures described in the
paper.

\section{The Optimized KL Filter}
\label{sec:KL}

\begin{figure}
\begin{center}
\includegraphics[width=0.98\columnwidth]{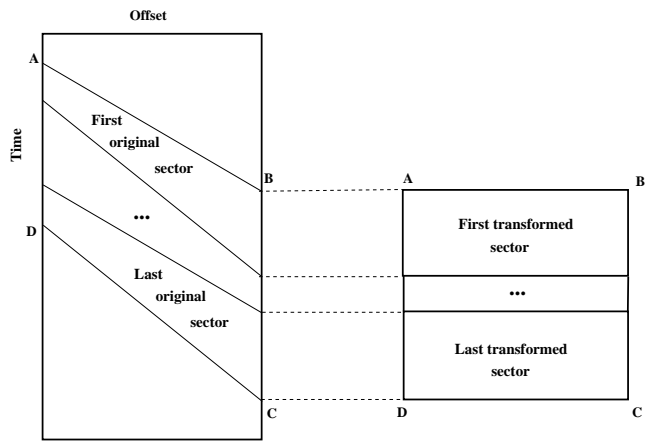}
\end{center}
\caption{Schematic diagram for demarcating the ground roll
on a seismic section and the corresponding rectangular sectors
obtained by applying a linear map.}
\label{fig:sector}

\end{figure}

As already mentioned, owing to its dispersive nature the ground-roll
noise appears in a seismic image as a typical fan-like coherent
structure.  This space-time localization of the ground roll allows us
to apply a sort of `surgical procedure' to suppress the noise, leaving
intact the uncontaminated region. To do that, we first pick lines to
demarcate the start and end of the ground roll  and, if
necessary, intermediate lines to demarcate different wavetrains, as
indicated schematically in Fig.~\ref{fig:sector}. In this figure we
have for simplicity used straight lines to demarcate the sectors but
more general alignment functions, such as segmented straight lines,
can also be chosen \cite{liu99,tyapkin2004}. To make  our
discussion as general as possible, let us  assume that we have a
set of $N$ parameters $\{\theta_i\}$, $i=1,...,N$, describing our
alignment functions. For instance, in Fig.~\ref{fig:sector} the
parameters $\{\theta_i\}$ would correspond to the coefficients of the
straight lines defining each sector.

Once the region contaminated by the ground roll has been demarcated, we
map each sector onto a horizontal rectangular region by shifting and
stretching along the time axis; see Fig.~\ref{fig:sector}. The data
points between the top and bottom lines in each sector is mapped into
the corresponding new rectangular domain, with the mapping being
carried out via a cubic convolution interpolation  technique
\cite{park83}. After this alignment procedure the ground roll events
will become approximately horizontal, favoring its decomposition in a
smaller space.  Since any given transformed sector has a rectangular
shape it can be represented by a matrix, which in turn can be
decomposed in empirical orthogonal modes (eigenimages) using the KL
transform.  The first few modes, which contain most of the ground
roll, are then subtracted to extract the coherent noise.  The
resulting data for each transformed sector is finally subjected to the
corresponding inverse mapping to compensate for the original forward
mapping. This leaves the uncontaminated data (lying outside the
demarcated sectors) unaffected by the whole filtering procedure.

The KL filter described above has indeed shown good performance in
suppressing source-generated noise from seismic data
\cite{liu99,tyapkin2004}. The method has however the drawback
that the region to be filtered must be picked by hand, which renders
the analysis somewhat subjective.  In order to overcome this
difficulty, it would be desirable to have a quantitative criterion
based on which one could decide what is the `best choice' for the
parameters $\{\theta_i\}$ describing the alignment functions. In what
follows, we propose an optimization procedure whereby the region to be
filtered can be selected automatically, once the generic form of the
alignment functions is prescribed.

Suppose we have chosen $l$ sectors to demarcate the different
wavetrains in the contaminated region of the original data, and let
$\{\theta_1,...,\theta_N\}$ be the set of parameters characterizing
the respective alignment functions that define these sectors. Let us
denote by $\tilde{A}_k$, $k=1,...,l$, the matrix representing the
$k$th transformed sector obtained from the linear mapping of the
respective original sector, as discussed above. For each 
transformed sector $\tilde{A}_k$ we then
compute its KL transform and calculate the {\it coherence index}
$CI_k$ for this sector, defined as the relative energy contained in
its first KL mode:
\begin{equation}
CI_k = \frac{\lambda_1^k}{\sum_{i=1}^{r_k} \lambda_i^k}, \label{eq:E1k}
\end{equation}
where $\lambda_i^k$ are the eigenvalues of the correlation matrix
$\tilde{\Gamma}_k=\tilde{A}_k\tilde{A}^t_k$ and $r_k$ is the rank of
$\tilde{A}_k$.  Such as defined above, $CI_k$ represents the relative
weight of the most coherent mode in the KL expansion of the
transformed sector $\tilde{A}_k$.  (A quantity analogous to our $CI$
is known in the oceanography literature as the similarity index
\cite{kim02}.)

Next we introduce an overall coherence index
$CI(\theta_1,...,\theta_N)$ for the entire demarcated region, defined
as the average coherence index of all sectors:
\begin{equation}
CI(\theta_1,...,\theta_N) = \frac{1}{l}{\sum_{k=1}^l  CI_k}. \label{eq:CI}
\end{equation}
As the name suggests, the coherence index $CI$ is a measure of the
amount of `coherent energy' contained in the chosen demarcated region
given by the parameters $\{\theta_i\}_{i=1}^N$. Thus, the higher $CI$
the larger the energy contained in the most coherent modes in that
region.  For the purpose of filtering coherent noise it is therefore
mostly favorable to pick the region with the largest possible $CI$. We
thus propose the following criterion to select the optimal region to
be filtered: vary the parameters $\{\theta_i\}$ over some appropriate
range and then choose the values $\theta_i^*$ that maximize the
coherence index $CI$, that is,
\begin{equation}
CI(\theta_1^*,...,\theta_N^*)= \max_{\{\theta_i\}}
\left[CI(\theta_1,...,\theta_N) \right].
\label{eq:CI*}
\end{equation}
Once we have selected the optimal region, given by the parameters
$\{\theta_i^*\}_{i=1}^N$, we then simply apply the KL filter to this
region as already discussed: we remove the first few eigenimages from
each transformed sector and inversely map the data back into the
original sectors, so as to obtain the final filtered image.  In the
next section we will apply our optimized KL filtering procedure to the
seismic data shown in Fig.~\ref{fig:sismo}.

\section{Results}
\label{sec:results}

Here we illustrate how our optimized KL filter works by applying it to
the seismic data shown in Fig.~\ref{fig:sismo}.  In this case, it
suffices to choose only one sector to demarcate the entire region
contaminated by the ground roll.  This means that we have to prescribe
only two alignment functions, corresponding to the uppermost and
lowermost straight lines (lines AB and CD, respectively) in
Fig.~\ref{fig:sector}. To reduce further the number of free parameters
in the problem, let us keep the leftmost point of the upper line
(point A in Fig.~\ref{fig:sector}) fixed to the origin, so that the
coordinates $(i_A,j_A)$ of point A are set to $(0,0)$, while allowing
the point $B$ to move freely up or down within certain range; see
below. Similarly, we shall keep the rightmost point of the lower line
(point C in Fig.~\ref{fig:sector}) pinned at a point $(i_C,j_C)$,
where $i_C=95$ and $j_C$ is chosen so that the entire ground roll
wavetrain is above this point. The other endpoint of the lower
demarcation line (point $D$ in Fig.~\ref{fig:sector}) is allowed to
vary freely. With such restrictions, we are left with only two free
parameters, namely, the angles $\theta_1$ and $\theta_2$ that the
upper and lower demarcation lines make with the horizontal axis. So
reducing the dimensionality of our parameter space allows us to
visualize the coherence index $CI(\theta_1,\theta_2)$ as a 2D surface.
For the case in hand, it is more convenient however to express $CI$
not as a function of the angles $\theta_1$ and $\theta_2$ but in terms
of two other new parameters introduced below.

Let the coordinates of point $B$, which defines the right endpoint of
the upper demarcation line in Fig.~\ref{fig:sector}, be given by
$(i_B,j_B)$, where $i_B=95$. In our optimization procedure we let
point $B$ move along the right edge of the seismic section by allowing
the coordinate $j_B$ to vary from a minimum value $j_{B_{min}}$ to a
maximum value $j_{B_{max}}$, so that we can write
\begin{equation}
j_B=j_{B_{min}}+k\Delta_B, \quad k=0,1,...,N_B
\label{eq:jb}
\end{equation}
where $N_B$ is the number of intermediate sampling points between
$j_{B_{min}}$ and $j_{B_{max}}$, and
$\Delta_B=(j_{B_{max}}-j_{B_{min}})/N_B$.  Similarly, for the
coordinates $(i_D,j_D)$ of point $D$ in Fig.~\ref{fig:sector}, which
is the moving endpoint of the lower straight line, we have $i_D=0$ and
\begin{equation}
j_D=j_{D_{min}}+l\Delta_D, \quad l=0,1,...,N_D,
\label{eq:jd}
\end{equation}
where $N_D$ is the number of  sampling points between $j_{D_{min}}$ and
$j_{D_{max}}$, and $\Delta_D=(j_{D_{max}}-j_{D_{min}})/N_D$.

\begin{figure}[h]
\begin{center}
\resizebox{70mm}{70mm}{ \includegraphics{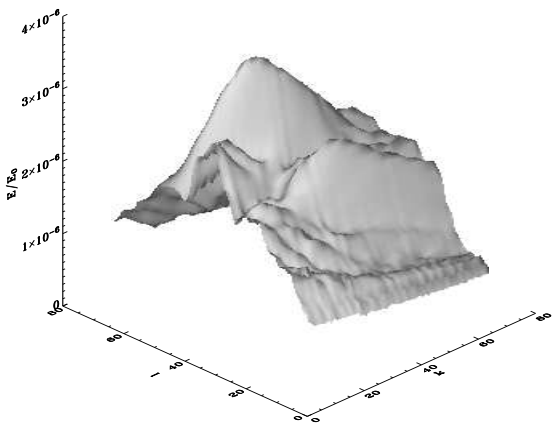}}
\caption{The coherence index $CI$  as a function of the indices $k$ and $l$
 that define the demarcation lines; see text.}
\label{fig:energ}
\end{center}
\end{figure}

\begin{figure}

%\vspace{-0.7cm}
\resizebox{42mm}{50mm}{%
\includegraphics*{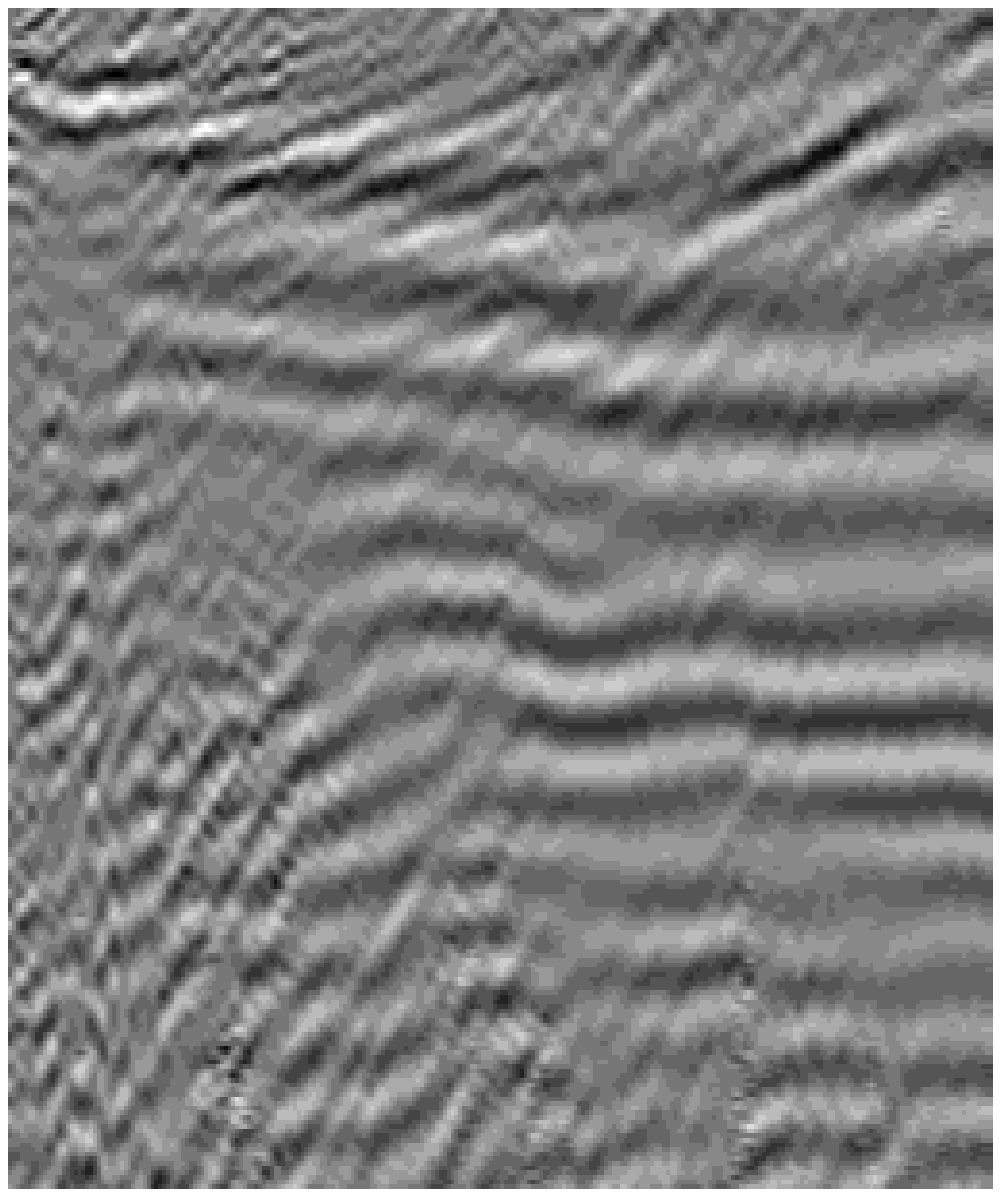}}
%\quad
\resizebox{42mm}{50mm}{%
\includegraphics{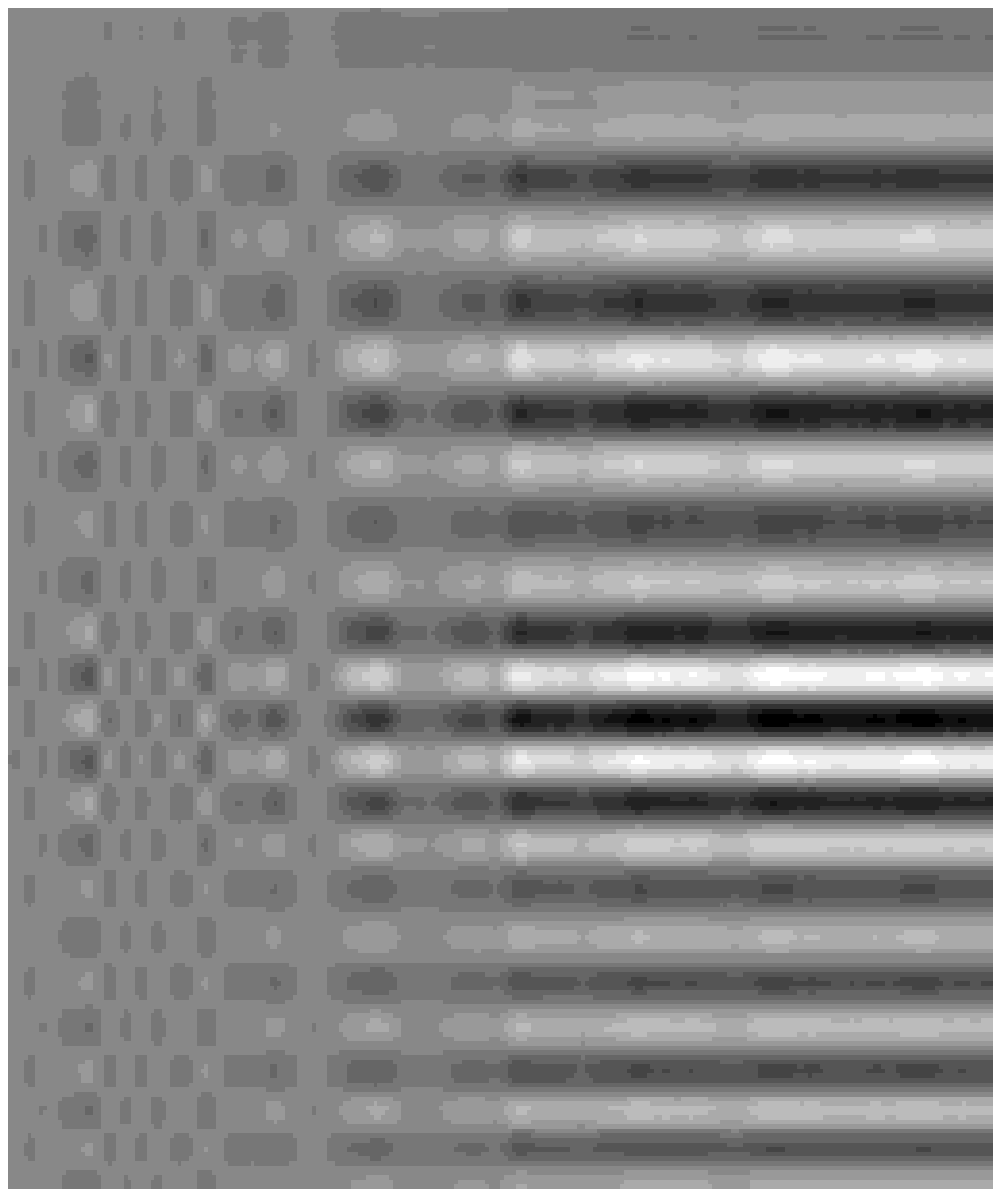}}
\resizebox{42mm}{50mm}{%
\includegraphics{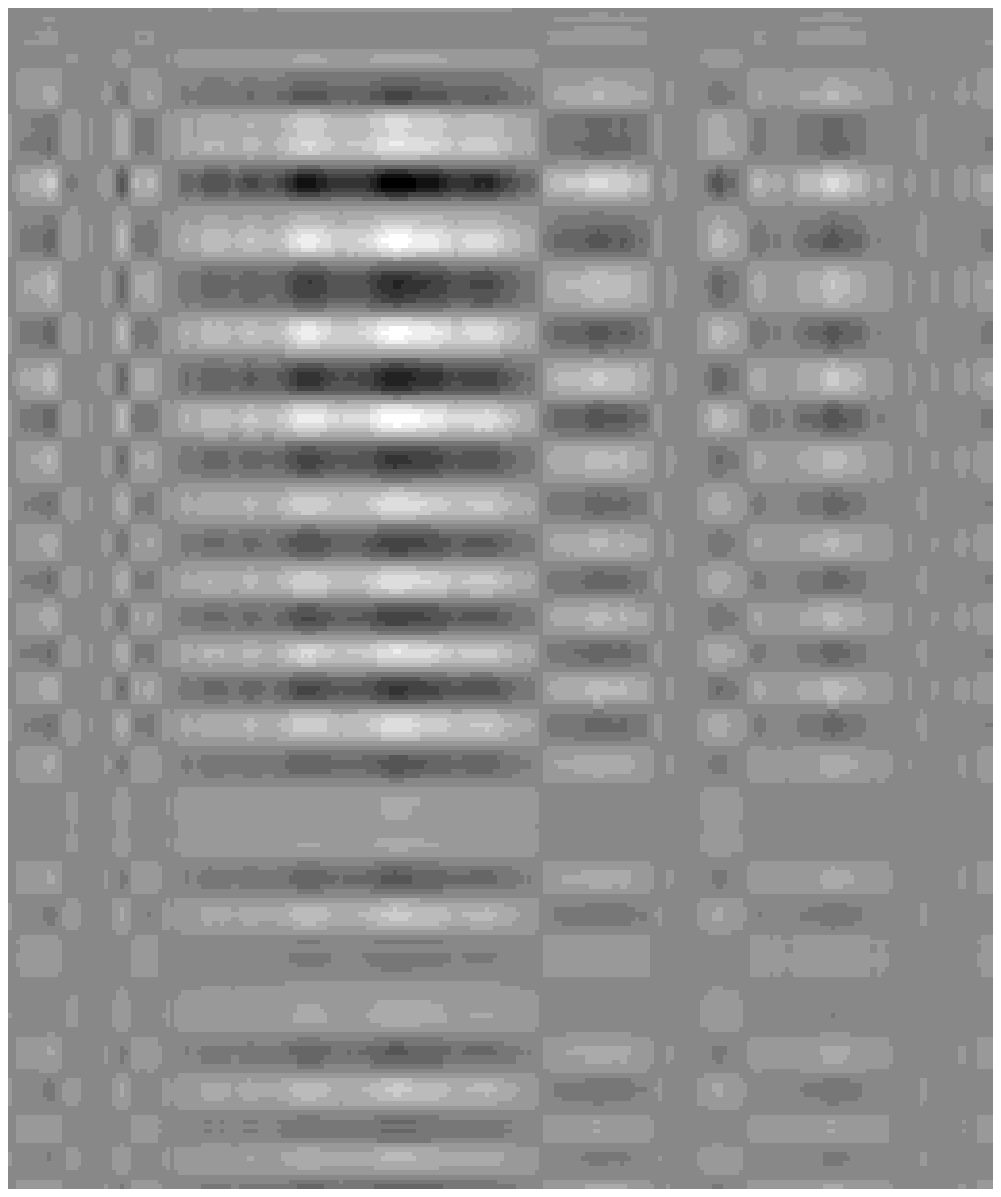}}%
\resizebox{42mm}{50mm}{%
\includegraphics{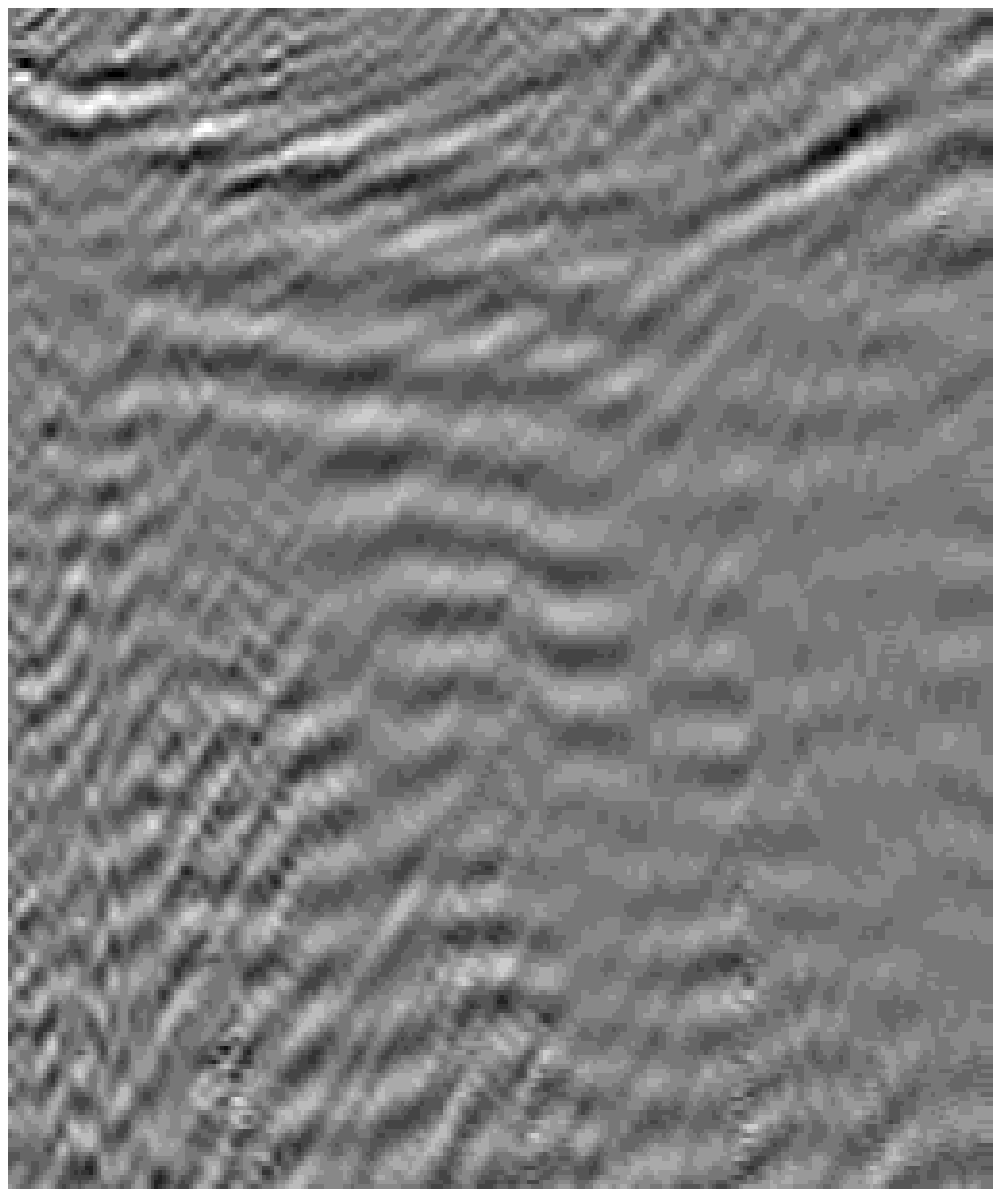}}%
\caption{ a) The selected region in the new domain;
b) its first eigenimage; c) the second eigenimage; and d) the result
after subtracting the first eigenimage.}

\label{fig:transf}

\end{figure}

\begin{figure}[h]
\begin{center}
\includegraphics[width=0.5\columnwidth]{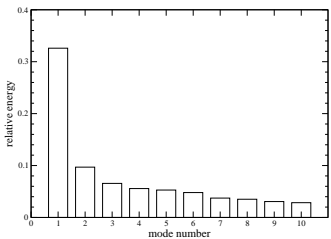}
\caption{The  relative energy of the first 10 KL modes of the region
shown in Fig.~\ref{fig:transf}a.}
\label{fig:modes}
\end{center}
                 
\end{figure}

For each choice of $k$ and $l$ in (\ref{eq:jb}) and (\ref{eq:jd}), we
apply the procedure described in the previous section and obtain the
coherence index $CI(k,l)$ of the corresponding region.  In
Fig.~\ref{fig:energ} we show the energy surface $CI(k,l)$, for the
case in which $j_{B_{min}}=280$, $j_{B_{max}}=600$, $j_C=864$,
$j_{D_{min}}=0$, $j_{D_{max}}=576$, and $N_B=N_D=64$.  We see in this
figure that $CI$ possesses a sharp peak, thus showing that this
criterion is indeed quite discriminating with respect to the
positioning of the lines demarcating the region contaminated by the
ground roll.  The global maximum of $CI$ in Fig.~\ref{fig:energ} is
located at $k=42$ and $l=24$, and in Fig.~\ref{fig:transf}a we show the
transformed sector obtained from the linear mapping of this optimal
region.  In this figure one clearly sees that the ground roll
wavetrains appear mostly as horizontal events.  In
Fig.~\ref{fig:transf}b we present the first eigenimage of the data
shown in Fig.~\ref{fig:transf}a, which corresponds to about 33\% of
the total energy of the image in Fig.~\ref{fig:transf}a, as can be
seen in Fig.~\ref{fig:modes} where we plot the relative energy $E_i$
captured by the first 10 eigenimages.  The second eigenimage, shown
in Fig.~\ref{fig:transf}c, captures about 10\% of the total energy,
with each successively higher mode contributing successively less to the
total energy; see Fig.~\ref{fig:modes}. In Fig.~\ref{fig:transf}d we
give the result of removing the first KL mode (Fig.~\ref{fig:transf}b)
from Fig.~\ref{fig:transf}a.  It is clear in Fig.~\ref{fig:transf}d
that by removing only the first eigenimage the main horizontal events
(corresponding to the ground roll) have already been greatly
suppressed.

\begin{figure}[h]
\begin{center}
\includegraphics[width=0.5\columnwidth]{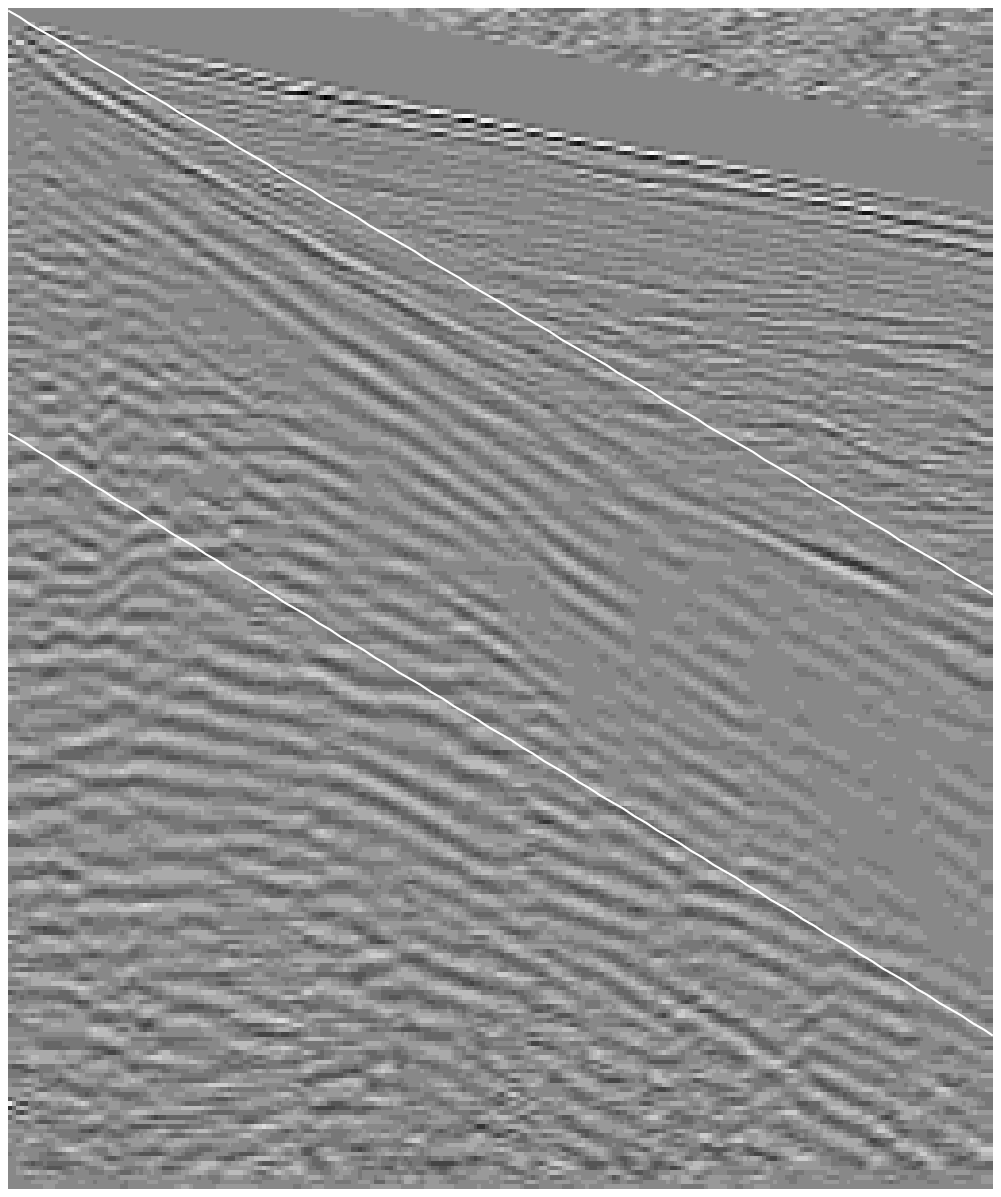}
\includegraphics[width=0.5\columnwidth]{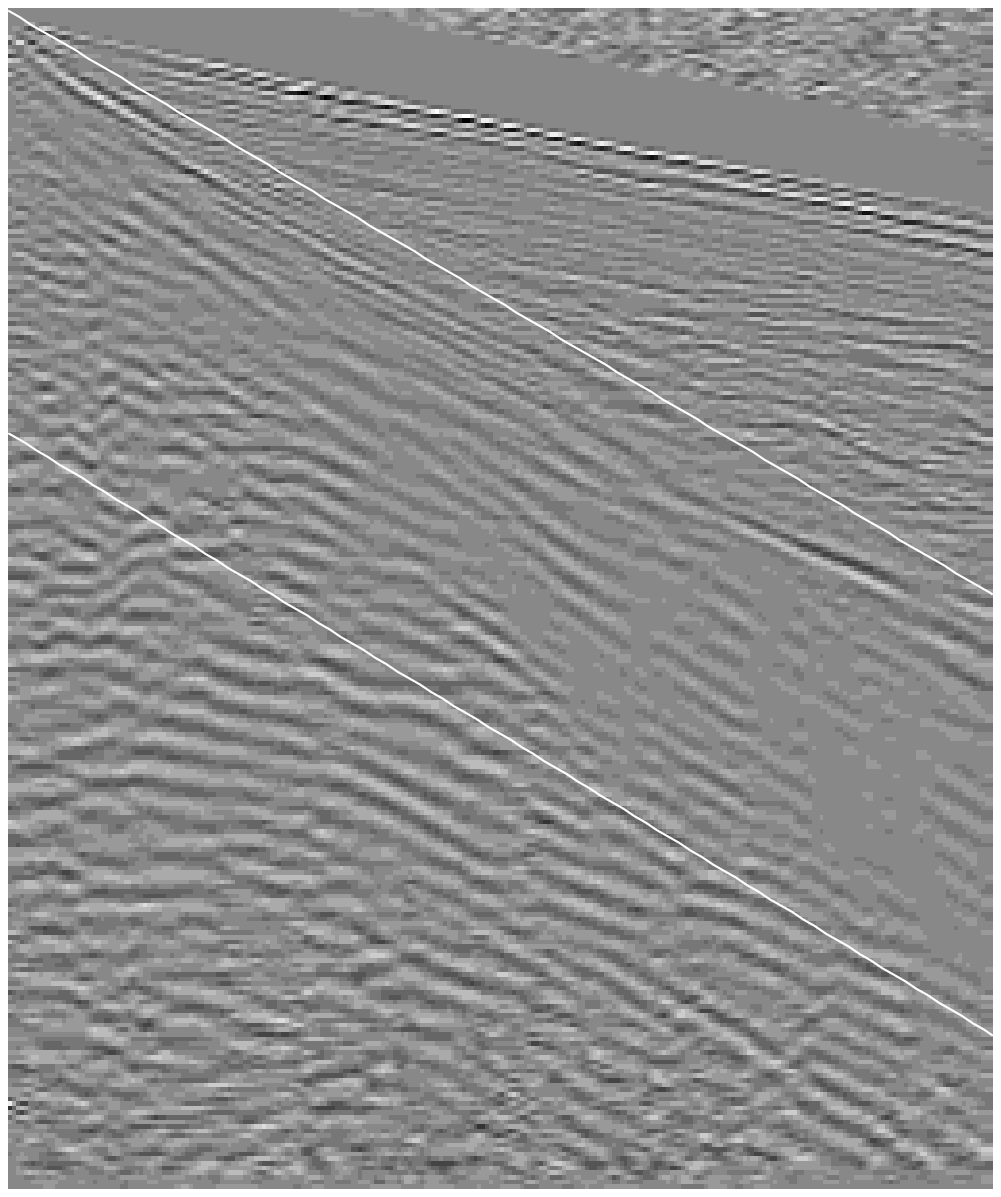}
\caption{a) The  filtered seismic section after removing 
the first
eigenimage  of the select region  shown in Fig.~\ref{fig:transf}a.
In b) we show the result after removing the first two eigenimages.}
\label{fig:sismoenerg}
\end{center}
\end{figure}

Performing the inverse mapping of the image shown in
Fig.~\ref{fig:transf}c yields the data seen in the region between the
two white lines in Fig.~\ref{fig:sismoenerg}a, which shows the final
filtered image for this case (i.e., after removing the first KL mode
from the transformed region).  We see that the ground roll inside the
demarcated region in Fig.~\ref{fig:sismoenerg}a has been considerably
suppressed, while the uncontaminated signal (lying outside the marked
region) has not been affected at all by the filtering procedure. If
one wishes to filter further the ground roll noise one may subtract
successively higher modes.  For example, in Fig.~\ref{fig:sismoenerg}b
we show the filtered image after we also subtract the second
eigenimage.  One sees that there is some minor improvement, but
removing additional modes is not recommended for it starts to degrade
relevant signal as well.

\section{Conclusions}
\label{sec:conclu}

An optimized filter based on the Karhunen--Lo\'eve transform has been
constructed for processing seismic data contaminated with coherent
noise (ground roll).  A great advantage of the KL filter lies in its
local nature, meaning that only the contaminated region of the seismic
record is processed by the filter, which allows the ground roll to be
removed without distorting most of the reflection signal. Another
advantage is that it is an adaptative method in the sense the the
signal is decomposed in an empirical basis obtained from the data
itself. We have improved considerably the KL filter by introducing an
optimization procedure whereby the ground roll region is selected so
as to maximize an appropriately defined coherence index $CI$.  
We emphasize that our method, require as input, only the generic
alignment functions to be used in the optimization procedure as well as
the number of eigenimages to be removed from the selected region.
These may vary depending on the specific application at hand.
However, once these choices are made, the filtering task can proceed
in the computer in an automated way.

Although our main motivation here has been suppressing coherent noise
from seismic data, our method is by no means restricted to geophysical
applications. In fact, we believe that the method may prove useful in
other problems in physics that require localizing coherent structures
in an automated and more refined way. We are currently exploring
further such possibilities.

\acknowledgments

Financial support from the Brazilian agencies CNPq and FINEP and from
the special research program \nobreak{CTPETRO} is acknowledged. We
thank L.~Lucena for many useful conversation and for providing us with
the data.

\appendix

\section{Relation between the KL transform and Proper 
Orthogonal Decomposition}
\label{sec:pod}

In dynamical systems the mathematical procedure akin to the KL
transform is called the proper orthogonal decomposition (POD).  In
this context, one may view each column vector $\vec{y}_j$ of the data
matrix $A$ as a set of $m$ measurements (real or numerical) of a given
physical variable $f(x,t)$ performed simultaneously at $m$ space
locations and at a certain time $t_j$, that is,
$y_{jk}=f(x=x_k,t=t_j)$, $k=1,...,m$. For example, in turbulent flows
the vectors $\vec{y}_i$ often represent measurements of the fluid
velocity at $m$ points in space at a given time $i$. The data matrix
$A$ thus corresponds to an ensemble $\{\vec{y}_j\}$ of $n$ such
vectors, representing  a sequence of $m$ measurements over $n$
instants of time.  In POD one is usually concerned with finding a
low-dimensional approximate description of the high-dimensional
dynamical process at hand. This is done by finding an `optimal' basis
in which to expand (and then truncate) a typical vector $\vec{y}$ of
the data ensemble.  Such a basis is given by the eigenvectors of the
time-averaged autocorrelation matrix $R$, which is proportional to the
matrix $\Gamma$ define above:
\begin{equation}
R\equiv\left<\vec{y} \, \vec{y}^{\; t}\right>=\frac{1}{n}\sum_{i=1}^n
\vec{y}_i \, \vec{y}_i^{\, t}=\frac{1}{n} \Gamma . \label{eq:R}
\end{equation}
Hence the eigenvectors $\{\vec{u}_i\}$ of $\Gamma$ are also
eigenvectors of $R$.  In POD parlance the eigenvectors $\{\vec{u}_i\}$
are called {\em empirical eigenvectors} or {\em proper orthogonal
modes}.  In the continuous case, the corresponding eigenfunctions
$u_i(x)$ of the autocorrelation operator are known as {\it empirical
orthogonal functions} (EOF).

From (\ref{eq:A}), (\ref{eq:psi}) and (\ref{eq:U}), one can  easily
verify that
\begin{equation}
\Psi_{ij}=\vec{u}_i\cdot\vec{y}_j. \label{eq:psiij}
\end{equation}
We thus see that the columns of the KL transform $\Psi$ correspond to
the coordinates of the vectors $\vec{y}$ in the empirical basis:
\begin{equation}
\vec{y}_i=\sum_{k=1}^{m} \Psi_{ki}\vec{u}_k. \label{eq:POD}
\end{equation}
It is this expansion of any member of the ensemble in the empirical
basis that is called the proper orthogonal decomposition or empirical
orthogonal function expansion.  It now follows from (\ref{eq:POD})
that
\begin{equation}
\left<\vec{y}^{\; 2}\right>=\frac{1}{n}\sum_{i=1}^n\vec{y}_i^{\; 2}=\frac{1}{n}
\sum_i (\Psi\Psi^t)_{ii}=\frac{1}{n}\sum_i \lambda_i, \label{eq:Ey}
\end{equation}
where in the last equality we used the fact that
\begin{equation}
\Psi\Psi^t=U^t\Gamma U=\Lambda,
\end{equation}
where $\Lambda$ is the diagonal matrix $\Lambda={\rm
diag}(\lambda_1,...,\lambda_m)$.  Equation (\ref{eq:Ey}) thus suggests
that we can interpret the eigenvalue $\lambda_i$ as a measure of the
energy in the $i$th empirical orthogonal mode.  For example, in the
case of turbulent flows where the vector $\vec{y}_i$ contains velocity
measurements at time $i$, the left hand of (\ref{eq:Ey}) yields twice
the average kinetic energy per unit mass, so that
$\frac{1}{2}\lambda_i$ gives the kinetic energy in the $i$th empirical
orthogonal mode \cite{holmes96}. Similarly, in the case of seismic
data the vectors $\vec{y}_i$ represent amplitudes of the reflected
waves, and hence the quantity $\sum_{i=1}^n\vec{y}_i^{\; 2}
=\sum_{i=i}^n\lambda_i$ may be viewed as a measure of the total energy
of the data, thus justifying the definition given in (\ref{eq:E}).

The optimality of the KL expansion also has a nice physical and
geometrical interpretation, as follows. Suppose we write a vector
$\vec{y}$ in an arbitrary orthonormal basis $\{\vec{e}_i\}_{i=1}^m$:
\begin{equation}
\vec{y}=\sum_{i=1}^m a_i \vec{e}_i,
\end{equation}
where $a_i=\vec{e}_i\cdot\vec{y}$.  If we now wish to approximate
$\vec{y}$ by only its first $k<m$ components,
\begin{equation}
\vec{y}^{\, k} = \sum_{i=1}^k a_i \vec{e}_i,
\end{equation}
then the optimality of the KL expansion implies that the first $k$
proper orthogonal modes capture more energy (on average) that the
first $k$ modes of any other basis. More precisely, the mean square
distance $\left<|\vec{y}-\vec{y}^{\, k}|^2 \right>$ is minimum if we
use the empirical basis.

\section{Relation between the KL transform and Principal Component Analysis}
\label{sec:pca}

In statistical analysis of multivariate data, the KL transform is
known as principal component analysis (PCA). In this case, one views
the elements of a row vector $\vec{x}_i=(x_{i1},...,x_{in})$ of the
data matrix $A$ as being $n$ realizations of a random variable $X_i$,
so that the matrix $A$ itself corresponds to $n$ samples of a random
vector $\vec{X}$ with $m$ components: $\vec{X}=(X_1,...,X_m)^t$.  In
other words, the column vectors $\vec{y}_j$ correspond to the samples
of $\vec{X}$. If the rows of $A$ are centered, i.e., the variables
$X_i$ have zero mean, then the matrix $\Gamma$ is proportional to the
covariance matrix $S_X$ of $\vec{X}$ \footnote{The proper
normalization for the covariance matrix is often chosen to be
$\frac{1}{n-1}$ instead of $\frac{1}{n}$, but this is not relevant for
our discussion here.}:
\begin{equation}
({S_X})_{ij}\equiv\left<X_iX_j\right> =\frac{1}{n}\vec{x_i}\cdot\vec{x_j}
=\frac{1}{n}\Gamma_{ij},
\end{equation}
or alternatively in matrix notation
\begin{equation}
S_X\equiv\left<\vec{X} \, \vec{X}^{\; t}\right>=\frac{1}{n}\Gamma. \label{eq:S}
\end{equation}
[Note that the matrices $R$ and $S_X$ defined respectively in
(\ref{eq:R}) and (\ref{eq:S}) are essentially the same but have
different interpretations.]  In the PCA context, the diagonal elements
$\Gamma_{ii}$ of the matrix $\Gamma$ are thus proportional to the
variance of the variables $X_i$, whereas the off-diagonal elements
$\Gamma_{ij}$, $i\ne j$, are proportional to the covariance between
the variables $X_i$ and $X_j$.  Furthermore, the eigenvectors
$\vec{u}_i$ of $\Gamma$ correspond to the principal axis of the
covariance matrix $S_X$. The idea behind PCA is to introduce a new set
of $m$ variables $P_i$, each of which being a linear combination of
the original variables $X_i$, such that these new variables are
mutually uncorrelated.  This is accomplished by projecting the vector
$\vec{X}$ onto the principal directions of the covariance matrix. More
precisely, we define the {principal components} $P_i$, $i=1,...,m$, by
the following relation
\begin{equation}
P_i=\vec{X}\cdot\vec{u}_i=\sum_{j=1}^n u_{ij} X_j. \label{eq:Pi}
\end{equation}
In other words, the vector of principal components
$\vec{P}=(P_1,...,P_m)^t$ is obtained from a rotation of the original
vector $\vec{X}$:
\begin{equation}
\vec{P}=U^t\vec{X}. \label{eq:P}
\end{equation}
The covariance matrix $S_P$ of the principal components is then given by
\begin{equation}
S_P=\left<\vec{P} \, \vec{P}^{\, t}\right>=\left<U^t\vec{X} \, \vec{X}^{\, t} U\right>
=\frac{1}{n} U^t\Gamma U=\frac{1}{n}\Lambda,
\end{equation}
thus showing that 
\begin{equation}
\left<P_iP_j\right>=0, \quad {\rm for} \quad i\ne j,
\end{equation} 
as desired.  The first principal component $P_1$ then represents the
particular linear combination of the original variables $X_i$ (among
all possible such combinations that yield mutually uncorrelated
variables) that has the largest variance, with the second principal
component possessing the second largest variance, and so on.

From (\ref{eq:psi}) and (\ref{eq:P}) one sees that the elements of the
$i$th row of the KL transform $\Psi$ correspond to the $n$ samples or
{\it scores} of the $i$th principal component.  That is, if we denote
the sample vector of the $i$th principal component by
$\vec{p}_i=(p_{i1},...,p_{in})$, then
$p_{ij}=\vec{u}_i\cdot\vec{y}_j=\Psi_{ij}$.  For this reason in the
PCA context the KL transform $\Psi$ is called the matrix of scores.

%%%%%%%%%%%%%%%%%%%%%%%%%%%%%
%\bibliography{allrhm}

\end{document}